\documentclass{webofc}
\usepackage[varg]{txfonts}   
\begin{document}
\title{ Global and local polarization of $\Lambda$ hyperons across RHIC-BES energies}
%
%

\author{
        \firstname{Xiang-Yu}\lastname{ Wu}\inst{1}\fnsep\thanks{{Speaker: xiangyuwu@mails.ccnu.edu.cn}} 
        \and  
        \firstname{Cong}\lastname{ Yi}\inst{2} 
        \and
        \firstname{Guang-You}\lastname{ Qin}\inst{1}\fnsep\thanks{\email{guangyou.qin@ccnu.edu.cn}} 
        \and 
        \firstname{Shi}\lastname{ Pu}\inst{2} 
}  

\institute{Institute of Particle Physics and Key Laboratory of Quark and Lepton
Physics (MOE), Central China Normal University, Wuhan, Hubei, 430079 
\and 
Department of Modern Physics, University of Science and Technology
of China, Hefei, Anhui 230026
}

\abstract{%
   We report our recent study on the global and local polarization of $\Lambda$ hyperons in Au+Au collisions at RHIC-BES energies within the (3+1)-dimensional CLVisc hydrodynamics framework. We present our numerical results for the global polarization as the function of collision energies and the local polarization along the beam direction as functions of azimuthal angle in $20-50$\% centrality at $\sqrt{s_{NN}}$=7.7 GeV Au+Au collision energy. We have discussed the effects of initial conditions, Spin Hall effect and baryon diffusion.
   }
\maketitle
\section{Introduction}
\label{intro}
Recently, spin polarization of $\Lambda$ and $\overline{\Lambda}$ hyperons along the beam and out-of-plane directions measured in relativistic heavy-ion collisions \cite{STAR:2019erd} is found to be different with various phenomenological models, which can explain the experimental data of global polarization \cite{Liang:2004ph, STAR:2017ckg}.
Several out-of-equilibrium effects, including shear induced polarization (SIP) and Spin Hall effects (SHE) have been introduced and widely discussed \cite{Fu:2021pok, Becattini:2021iol,Yi:2021ryh,Alzhrani:2022dpi}. 
On the other hand, model uncertainties still exist. In this paper, we report our recent study on the sensitivity of spin polarization to initial conditions, SHE, and baryon diffusion current. 

\section{Theoretical framework}
\label{sec-1}
The averaged spin polarization vector for the spin-$1/2$ particles can be computed via the modified Cooper-Frye formula \cite{Becattini:2013fla,Fang:2016vpj}, $\mathcal{S}^\mu(\mathbf{p})=\int d \Sigma \cdot p \mathcal{J}_5^\mu(p, X) / [2 m \int d \Sigma \cdot \mathcal{N}(p, X)],$ 
where $\mathcal{J}_5^\mu(p, X)$ denotes the axial-charge current density  in phase space and $\mathcal{N}(p, X)$ is the particles' distribution function. 
Under the assumption of local thermal equilibrium in the quantum kinetic theory \cite{Hidaka:2017auj}, the total spin polarization vector $\mathcal{S}^\mu(\mathbf{p})$ can be induced by thermal vorticity, shear tensor, fluid acceleration, Spin Hall effect, and electric magnetic fields, i.e. $\mathcal{S}^\mu(\mathbf{p})=  \mathcal{S}_{\text {thermal }}^\mu(\mathbf{p})+\mathcal{S}_{\text {shear }}^\mu(\mathbf{p})+\mathcal{S}_{\text {accT }}^\mu(\mathbf{p}) 
 +\mathcal{S}_{\text {chemical }}^\mu(\mathbf{p})+\mathcal{S}_{\mathrm{EB}}^\mu(\mathbf{p})
$. The polarization induced by magnetic fields has been estimated via the anomalous magneto-hydrodynamics \cite{Peng:2023rjj} and will be neglected in the current studies. 
Polarization induced by other sources can be written in terms of hydrodynamics variables, e.g. the temperature, chemical potential and gradient of fluid velocity. The explicit expression for these $\mathcal{S}$ can be found in our previous works \cite{Yi:2021ryh,Wu:2022mkr}. In current study, we utilize the (3+1)-dimensional CLVisc hydrodynamics \cite{Wu:2021fjf} to simulate the evolution of quark gluon plasma (QGP) and provide the profiles of temperature and vorticity.
Later on, we take a Lorentz transformation to derive $\Lambda$ polarization to its own local rest frame,  maintaining consistency with experimental measurements. At last, the polarization along $y$ or $z$ directions, named $P^y,P^z$, as functions of azimuthal angle $\phi_p$, can be obtained through the integrating normalized $\mathcal{S}$ over rapidity $\eta$ and transverse momentum $p_T$. 

\begin{figure}[h]
\centering
\includegraphics[scale=0.35]{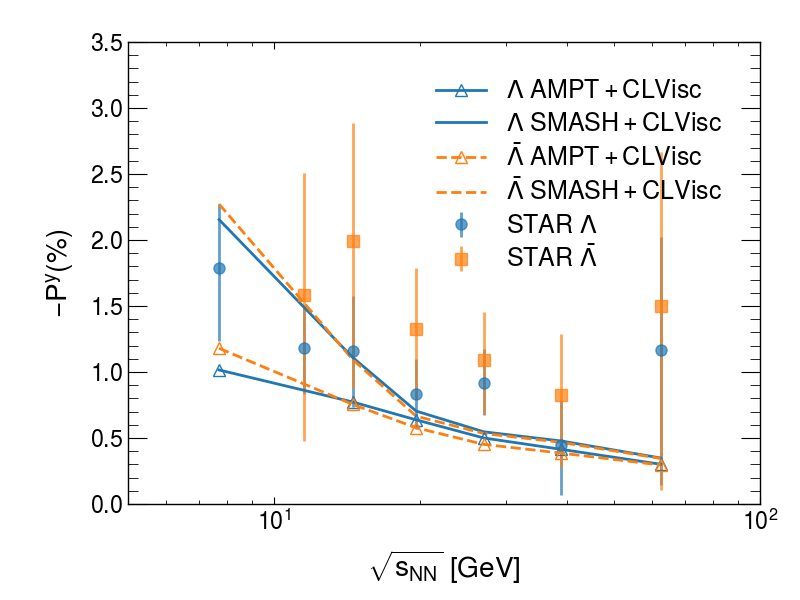}
\caption{
The collision energy dependence of global polarization $P^{y}$ for $\Lambda$ and $\bar{\Lambda}$ in 20-50\% centrality in Au+Au collisions. The experimental data are taken from STAR collaboration\cite{STAR:2021beb}, compared to the hydrodynamics calculation with AMPT and SMASH initial conditions. }
\label{fig-1}       
\end{figure}

\section{Numerical results}
\label{sec-2} 
We present the numerical results for the global polarization of $\Lambda$ and $\bar{\Lambda}$ hyperons along the out-of-reaction plane, $P^y$, at Au+Au collisions across RHIC-BES in Fig ~\ref{fig-1}. To compare with experiment data, we have chosen the mid-rapidity region $|\eta|$ < 1,  $p_T$ range [0.5,3.0] GeV and $20-50$\% collision centrality. The results simulated with both AMPT~\cite{Lin:2004en} and SMASH~\cite{SMASH:2016zqf} initial condition models agree  with the STAR data well within the error range.
We observe that the global polarization at low collision energy with the SMASH initial model has larger magnitude than those with the AMPT initial model, due to the finite nuclear effect. While, the spin polarization from both two initial conditions presents similar results at high collision energies. It means that global polarization is sensitive to the state of nuclear matter at the low collision energy. On the other hand, we also observe a slight difference between spin polarization of $\Lambda$ and $\bar{\Lambda}$ hyperons, when considering the effect of finite baryon chemical potential in our simulations.

\begin{figure}[h]
\centering
\includegraphics[scale=0.3]{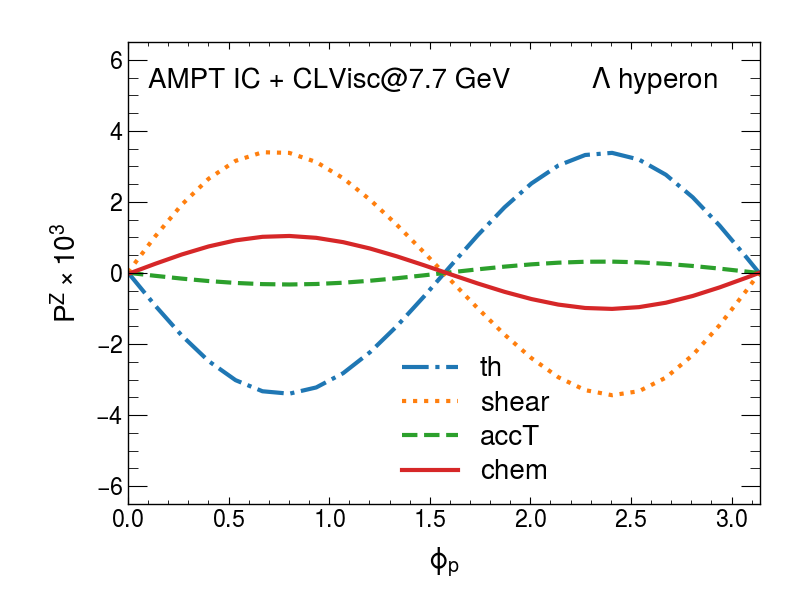}
\includegraphics[scale=0.3]{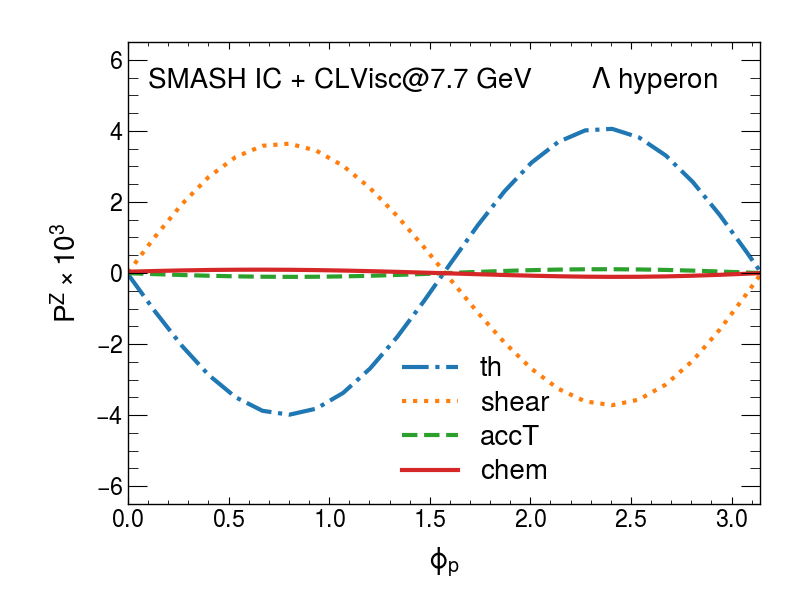}
\caption{The azimuth angle dependence of local longitudinal polarization of $\Lambda$ hyperon $P^Z$ in 20-50\% centrality in Au+Au collisions  at $\sqrt{s_{NN}}$ = 7.7 GeV for hydrodynamics simulations with AMPT initial condition (left) and SMASH initial condition (right). The lines with different styles represent the separate contributions from the thermal vorticity (blue, dash dot ), SIP (orange, dot), fluid acceleration (green, dash),  SHE (red, solid). }
\label{fig-2}       
\end{figure}

In Fig.~\ref{fig-2}, we plot the local spin polarization along the beam direction, $P^{z}$, contributed from different sources under the AMPT and SMASH initial conditions as a function of the azimuthal angle $\phi_p$ in $20-50$\% centrality at $\sqrt{s_{NN}}$=7.7 GeV Au+Au collision energy. 
The local spin polarizations along the beam direction, $P^{z}$, induced by SIP and SHE show the sin shape with respect to $\phi_p$, giving a correct tend compared to the STAR data at $\sqrt{s_{NN}}=200$ GeV \cite{STAR:2019erd}. Conversely, the contribution from the thermal vorticity and fluid acceleration give negative signs. The magnitude of $P^z$ induced by SIP are compared with those induced by thermal vorticity with two initial conditions. The polarization induced by SHE strongly depends on the initial condition and it is  negligible with the SMASH initial condition. 

\begin{figure}[h]
\centering
\includegraphics[scale=0.3]{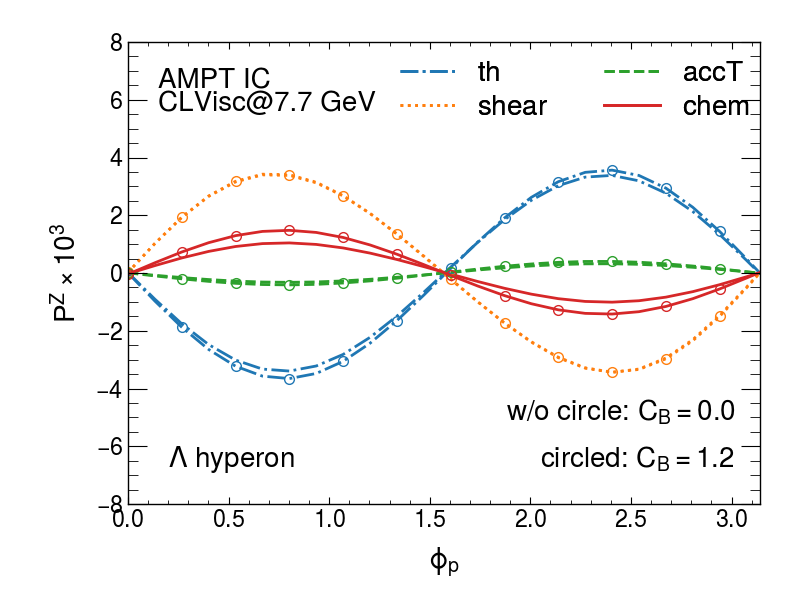}
\includegraphics[scale=0.3]{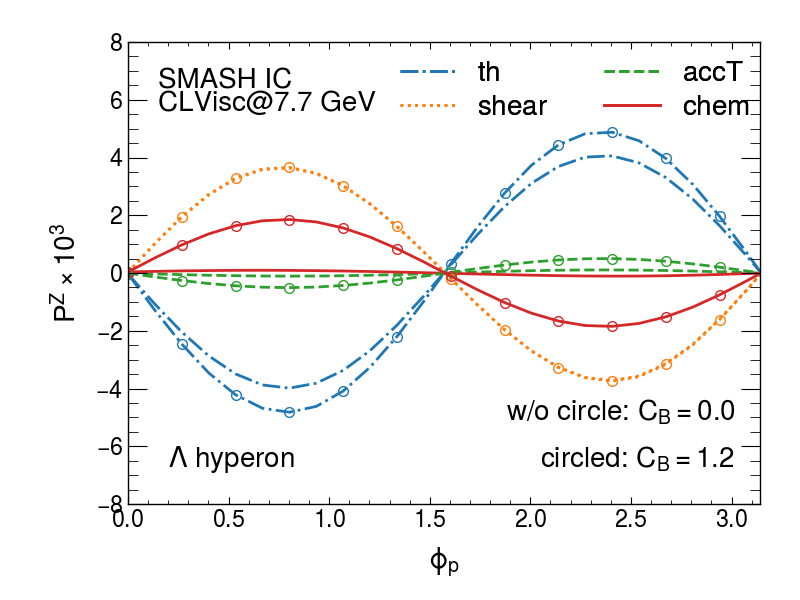}
\caption{The azimuth angle dependence of local longitudinal polarization of $\Lambda$ hyperon $P^Z$ in $20-50$\% centrality in Au+Au collisions  at $\sqrt{s_{NN}}$ = 7.7 GeV for hydrodynamics simulations with AMPT initial condition (left) and SMASH initial condition (right). The lines with and without open circle stand for baryon diffusion with $C_B$=1.2 and $C_B$=0.0, respectively. }
\label{fig-3}       
\end{figure}

At last, we study the effect of baryon diffusion to the $P^z$ in $20-50$\% centrality at $\sqrt{s_{NN}}$ = 7.7 GeV Au+Au collisions in Fig.~\ref{fig-3}.
We find that $P^z$ induced by thermal vorticity and SIP slightly increases when baryon diffusion coefficient $C_B$ increases. 
We also observe that with SMASH initial condition the $P^z$ induced by SHE enhances significantly with $C_B$ increasing.  

\section{Summary}

We have studied the dependence of global polarization of $\Lambda$ hyperons on the collisional energies and find that global polarization is sensitive to the initial condition, especially at low collision energies.
We also discuss local polarization $P^z$ as functions of azimuthal angle at $\sqrt{s_{NN}}= 7.7$  GeV Au+Au collisions. 
The $P^z$ induced by SHE has a positive contribution to the total one and is sensitive to initial conditions and baryon diffusion coefficients $C_B$. 
Our study indicates that there are still significant uncertainties in the numerical framework for calculating the spin polarization. The further systematical studies are needed.

\section*{Acknowledgements:} 
This work is supported in part by the National Key Research and Development Program of China under Contract No. 2022YFA1605500, by the Chinese Academy of Sciences (CAS)  under Grants No. YSBR-088 and by
NSFC under Grant Nos. 12075235, 12225503, 11890710, 11890711, 11935007, 12175122, 2021-867, 11221504, 11861131009 and 11890714. 

%
%
%

\end{document}